\documentstyle[11pt,Cargesepasp,twoside,epsf]{article}
\def\edcomment#1{\iffalse\marginpar{\raggedright\sl#1\/}\else\relax\fi}
\reversemarginpar

\begin{document}
\title{The Lithium Depletion Boundary as a Clock and Thermometer}
 \author{R.D. Jeffries \& T. Naylor}
\affil{Department of Physics, Keele University, Keele, Staffordshire,
 ST5 5BG}

\begin{abstract}
We take a critical look at the lithium depletion boundary (LDB) technique
that has recently been used to derive the ages of open clusters. We
identify the sources of experimental and systematic error and show that
the probable errors are larger by approximately a factor two than
presently claimed in the literature. We then use the Pleiades LDB age
and photometry in combination with evolutionary models to define empirical
colour-$T_{\rm eff}$ relations that can be applied to younger
clusters. We find that these relationships {\em do not} produce
model isochrones that match the younger cluster data. We propose that
this is due either to systematic problems in the evolutionary models or
an age (gravity) sensitivity in the colour-$T_{\rm eff}$ relation which
is not present in published atmospheric models.

\end{abstract}

\section{Introduction}
Lithium is an ephemeral element in the atmospheres of low mass pre main
sequence (PMS) stars. It burns swiftly
in $p,\alpha$ reactions once the core temperature, $T_{c}$, approaches
$3\times 10^{6}$\,K, with a reaction rate approximately proportional to
$T_{c}^{20}$ (Ushomirsky et al. 1998). Because PMS stars less massive
than 0.35M$_{\sun}$ are always fully convective, and because timescales
for convective mixing are much shorter than evolutionary timescales, the
surface abundance of Li is also rapidly depleted when the core reaches
this Li-burning temperature. Thus, for stars of a given mass, total Li
depletion occurs over a relatively narrow range of effective
temperatures and luminosities.  The time taken for $T_{c}$ to reach the
Li ignition point is sensitively dependent on mass and to a lesser
extent on adopted equations of state and atmospheric boundary
conditions. As a result, if one observes a group of co-eval stars, 
the mass, and hence luminosity and temperature,
at which Li is observed to be depleted from its initial value, offers a
potentially precise determination of stellar age.

Analytical (Bildsten et al. 1997; Ushomirsky et al. 1998) and numerical
(D'Antona \& Mazzitelli 1997 (DAM97); Chabrier \& Baraffe 1997 (CB97);
Burrows et al. 1997 (B97)) treatments of this process have
appeared. These models confirm that the ``lithium depletion boundary''
(LDB) is an independent age diagnostic for clusters of stars. The
technique is likely to be more accurate than the better known method of
fitting the upper main sequence turn-off. This is partly because the
uncertain physics in high mass stellar evolution models (convective
core overshoot, semi-convection, mass loss, rotational mixing) has
greater effects on the observable parameters than the debatable physics
in low mass models (equation of state, convection treatment and
atmospheres), and partly because there are usually only a few high mass
stars (with uncertain duplicity) with which to fit isochrones, whereas
there are many low mass stars around the LDB.  The LDB method is also
potentially superior to fitting isochrones to PMS stars as they descend
towards the ZAMS. The main hazard with this technique is the extremely
uncertain conversions between colours and effective temperatures for
$T_{\rm eff}<4500$\,K.

In this paper we provide an independent analysis of the likely
experimental and systematic errors in the LDB technique. We also show
that the LDB ages can be used in combination with observational data to
yield the effective temperatures of low-mass stars and use this as an
additional consistency check for the current generation of evolutionary
models.

\section{The LDB Error Budget}

Our baseline assumption is that the age of a cluster is found by
locating the LDB in a colour-magnitude diagram (see Fig.1). We convert
the apparent magnitude of the LDB into an absolute magnitude and then
into $L_{\rm bol}$ using an empirical bolometric correction at the
colour of the LDB. We then use models for Li depletion in cool stars to
translate $L_{\rm bol}$ at the LDB into an age. Note, that we {\em do
not} use the $T_{\rm eff}$-colour relationships or bolometric
corrections provided by some evolutionary models as these would instil
unquantifiable errors in the procedure and in several cases,
particularly in the optical, model colours have already been shown to
be systematically in error (Baraffe et al. 1998).

Experimental errors in the LDB age can be ascribed 
to uncertain placement of the LDB in a
colour-magnitude diagram due to sparse sampling in the Li measurements,
uncertain photometric calibrations and
uncertainties in the distance and reddening of the cluster in
question. Systematic errors arise from the chosen empirical bolometric
correction-colour relation, whether one assumes the LDB defines the
point at which say 90\% or 99\% of Li has been depleted and the choice
of evolutionary model which defines the $L_{\rm bol}$-age relation at
the LDB. The first two of these systematics will alter the ages of all
clusters in qualitatively the same way, whereas the latter systematic
may cause individual  clusters to be older or younger, as the $L_{\rm
bol}$-age relations cross for differing evolution models. There may of
course be additional systematics due to common assumptions made by all
the current generation of models.

\begin{figure}
\plotone{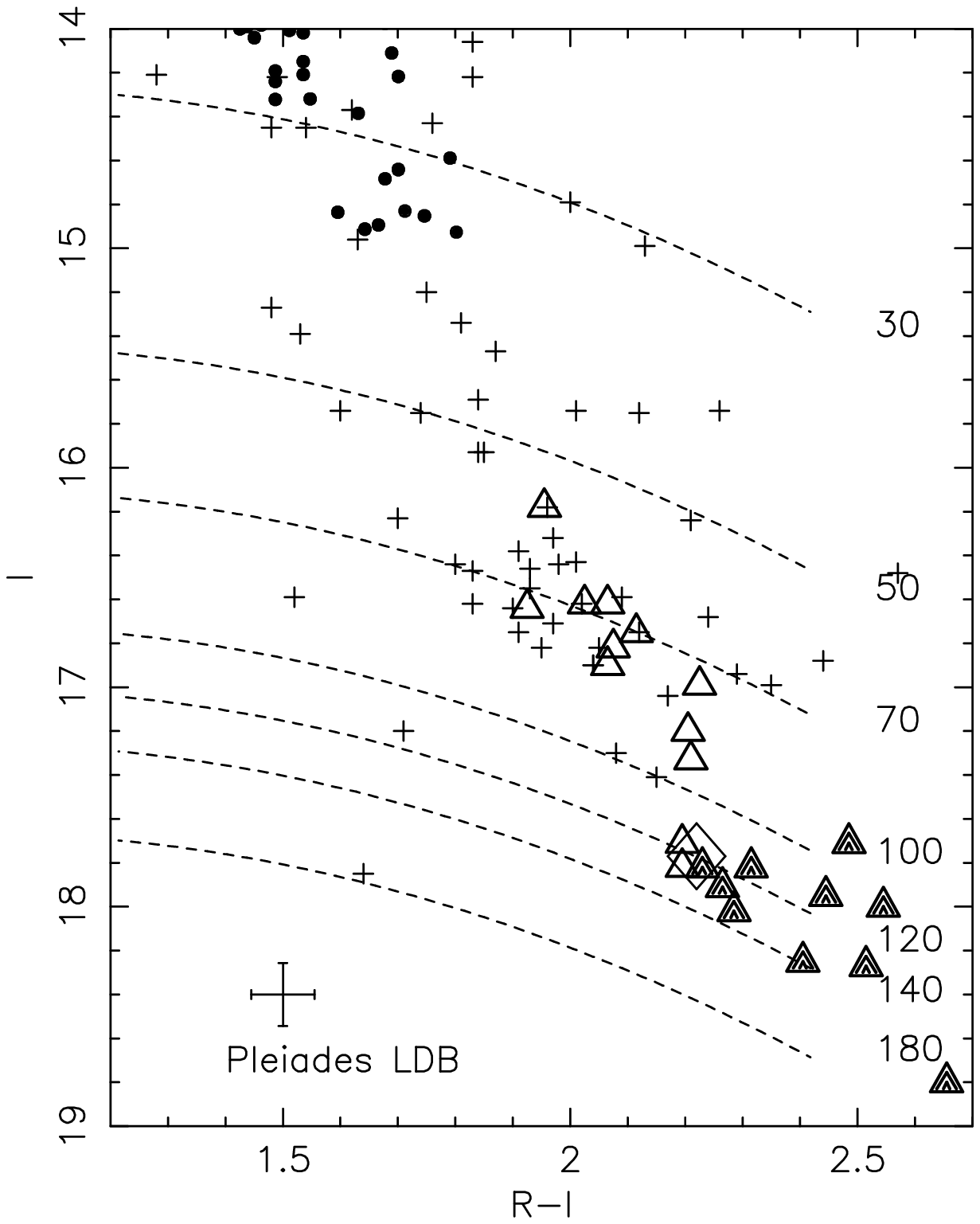}
\caption{An $I$ versus $R-I$ plot for the Pleiades. Triangles are stars
with detailed spectroscopy of the Li\,{\sc i} 6708\AA\ line. Filled triangles
have a Li\,{\sc i}\,6708\AA\ EW above 0.3\AA, while open triangles show no Li. 
Spots are proper motion members with
photoelectric photometry, crosses are
proper motion members with mainly photographic photometry.
Dashed lines are
isochrones of 99\% Li depletion using the models of CB97,
labelled in Myr. The diamond
marks the position we have chosen for the LDB and the cross represents the other random
errors (distance modulus, photometric calibration etc.) that make the
LDB age determination uncertain.}
\end{figure}

\begin{figure}
\plotone{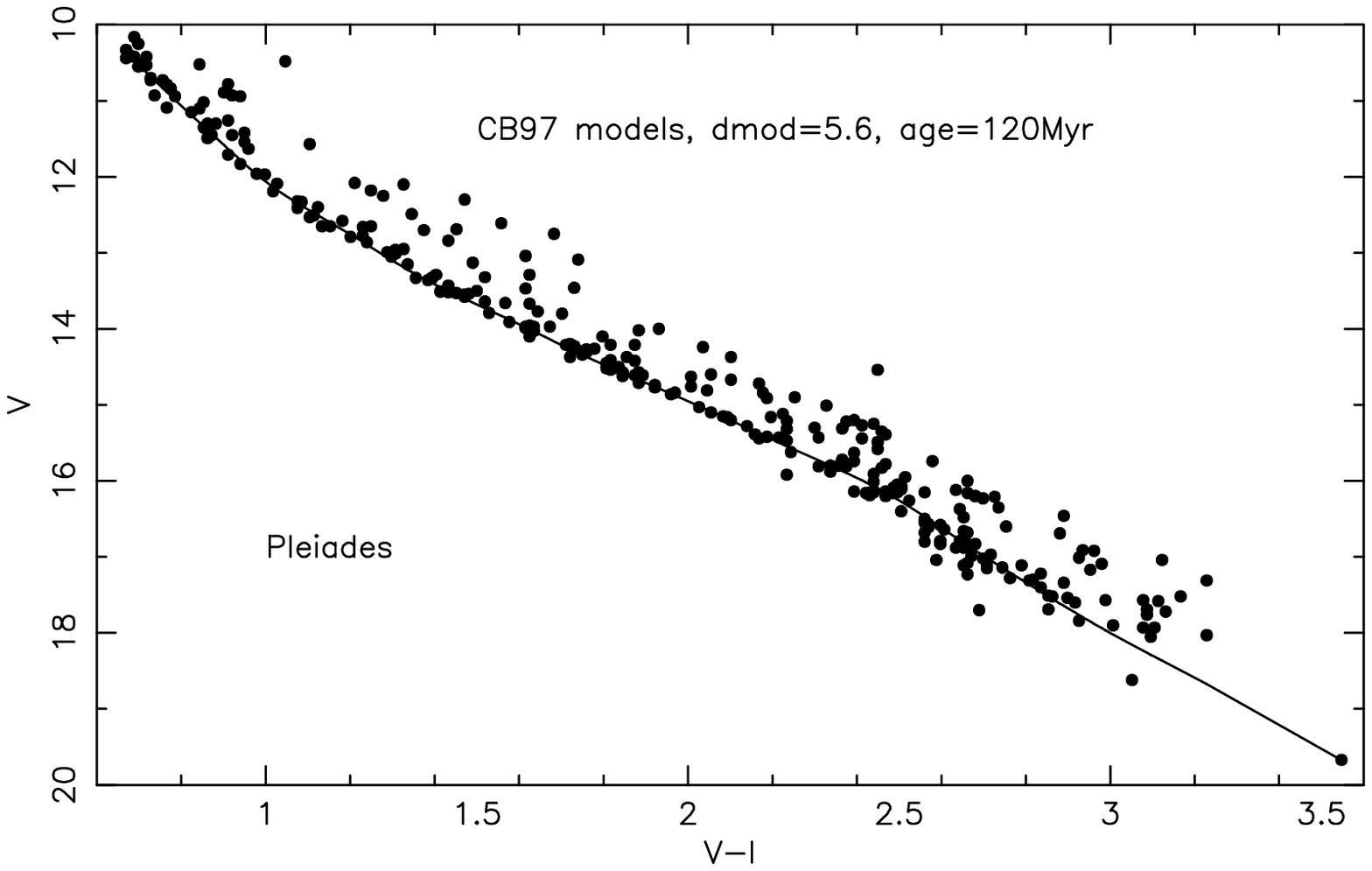}
\caption{An isochrone fit to the Pleiades at 120\,Myr, obtained
by tuning the $T_{\rm eff}-(V-I)$ relationship.}
\end{figure}

\begin{figure}
\plotone{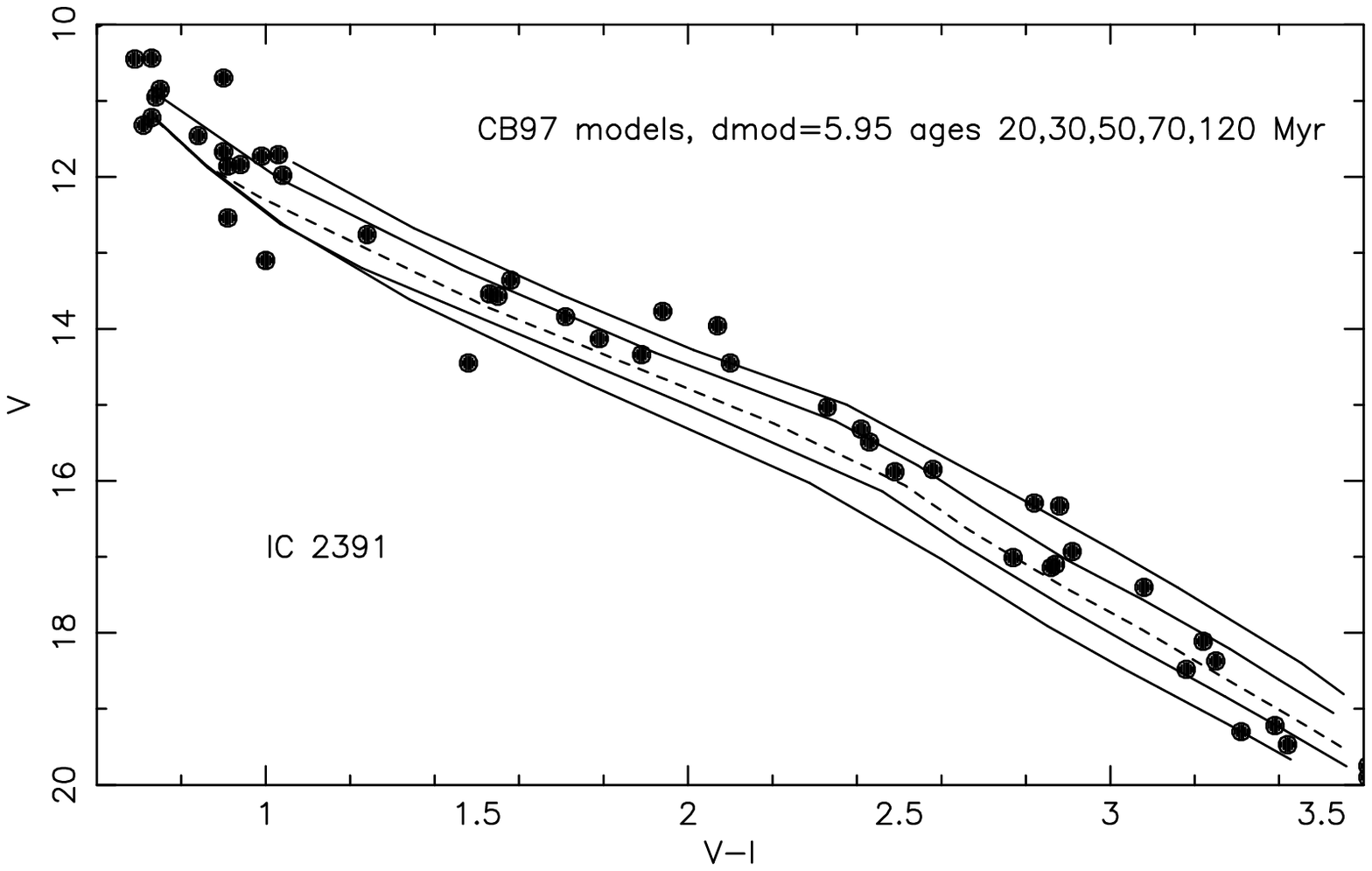}
\caption{Isochrones fitted to the IC 2391 data, assuming the
same $T_{\rm eff}-(V-I)$ relationship that worked for the Pleiades
data. The LDB age for IC2391 is indicated by the dashed isochrone.}
\end{figure}

We have investigated these uncertainties, focussing on clusters with 
LDB ages defined in the $I$ vs $R-I$ diagram, and using the
theoretical models of B97, CB97 and DAM97. We find 
that for clusters with
data quality similar to that available for the Pleiades (Fig.1), the experimental
error in age rises from $\pm9\%$ at $\sim25$\,Myr, to $\pm16\%$ at
200\,Myr. These errors are dominated by uncertain placement of the LDB
(typically $\pm 0.15$ mag in $I$ and $\pm 0.05$ in colour) along with
distance modulus errors and uncertainties in the $I$ photometric
calibration (both around $\pm0.1$ mag). Obtaining Li data for more
points around the LDB, distinguishing binaries from single stars and
tightening up the photometric calibration could reduce these errors
considerably.  The increase in error with age is due to the propinquity
of the Li depletion isochrones at older ages.  The systematic errors
are more constant, ranging between $7\%$ and $11\%$ over a
similar age range.  The systematics are due mainly to the choice of
model and whether there is any age dependence in the bolometric
corrections as a function of colour, which are usually derived from
mature stars. The influence of whether Li is
depleted by 90\% or 99\% at the LDB (say for Li 6708\AA\ EWs $<0.3$\AA)
only alters ages by $\pm3$\%.

\begin{table}
\caption{LDB Ages and errors for young clusters}
\begin{tabular}{crrr}
\hline
\hline
&&&\\
Cluster & LDB Age (Myr) & Experimental Error & Systematic Error \\
&&&\\
\hline
&&&\\
Pleiades & 122 & $+20/-17$ & $\pm11$ \\
&&&\\
Alpha Per & 85 & $+12/-10$ & $\pm8$ \\
&&&\\
IC 2391 & 48   & $\pm5$ & $\pm3$ \\
&&&\\
\hline
\end{tabular}
\end{table}

Table~1 and Figure 1 present example results of our analysis for the
Pleiades, Alpha Per and IC 2391 clusters using Li spectra, $I$ and
$R-I$ data from the literature. The ages result from the average of
using the CB97, B97 and DAM97 models.  These ages
are similar to those previously estimated (Stauffer et al. 1998,
1999; Barrado y Navascu\'{e}s et al. 1999), but the estimated errors are
larger by a factor of two. This does not alter the main conclusions of
this set of papers -- that the
LDB ages are older (by a factor 1.6 or so) 
than high mass turn-off ages with no core overshoot.
However, the larger errors certainly obscure whether the amount of core
overshoot required to bring turn-off and LDB ages into agreement
might be mass dependent.

\section{The effective temperature-colour relationship}

The $T_{\rm eff}$-colour relationship in cool ($<4500$\,K) stars has long
been a source of considerable uncertainty. If one {\em knew} the ages
of clusters, then their photometric data provide an empirical
isochrone from which a $T_{\rm eff}$-colour relationship can be
derived, such that a theoretical isochrone at the same age will match
the same data. If one further assumes the $T_{\rm eff}$-colour
relationship is age-independent, then one could produce
isochrones in colour-magnitude diagrams which should match other
clusters at their LDB ages. 

We have performed these tests using 
the CB97 and DAM97
models and with $V$ vs $V-I$ and $I$ vs $R-I$ diagrams, 
with very similar results. Figure 2 shows an example of the
empirical isochrone defined by 
the Pleiades $V$ vs $V-I$ diagram obtained by choosing a
$T_{\rm eff}$-colour relation to yield a good match at the LDB age and assuming
a distance modulus of 5.6. 
Figure 3 shows the result of using this relationship to try and fit
isochrones to the available IC 2391 data. The indicated
isochronal age ($\simeq30$\,Myr) is younger than the LDB age and may be in
keeping with the more traditional nuclear turn-off age. There
is some indication that the ages converge for the coolest stars in the
sample. However, tests using the available
$R-I$ data in the Pleiades and IC 2391 for even cooler stars show that
the discrepancy still exists there, so we attribute the convergence
perhaps to poor photometric calibration at the most extreme colours. 
Changing the
age (or distance modulus) of the Pleiades (or IC 2391) within their
errors makes no difference to
this conclusion. Using the Hipparcos Pleiades distance modulus of 5.3 makes the
Pleiades LDB age older by $\sim 25$\,Myr and increases the discrepancy
between isochronal and LDB ages for IC 2391. 

\section{Conclusions}

Could the discrepancy between isochronal and LDB ages, 
assuming a universal $T_{\rm eff}$-colour relationship, mean
that there are systematic problems in the current generation of low-mass
evolution models? Or, could the $T_{\rm eff}$-colour relationship change
sufficiently between 50\,Myr and 125\,Myr to explain our results?
Given the discrepancies between LDB and isochronal ages, this also begs the question as
to how accurate isochronal ages derived in even younger clusters and
associations could possibly be? If the LDB ages are correct and our
empirical isochrones are not, then the implication is that low-mass
objects in very young star forming regions and associations could be
twice as old as inferred from current evolutionary models combined with
similar $T_{\rm eff}$-colour relations.

The models of Baraffe et al. (1998) suggest that any age dependence of
the $T_{\rm eff}$-colour relationship is small in this age
range. However, the reliability of the optical colours, particularly
$V-I$ and $R-I$, in these models has been called into question by the
authors themselves, so any conclusions based on these is perhaps
premature. Unfortunately there is presently insufficient near IR data
in young clusters, especially IC 2391, to do a similar test in a
spectral region where the model colours are likely to be more realistic.

\end{document}